\documentclass{aa}  

\usepackage{graphicx}
\usepackage[varg]{txfonts}
\usepackage{appendix}
\bibpunct{(}{)}{;}{a}{}{,}
\usepackage{color}

\begin{document}

   \title{Spatial metallicity variations of mono-temperature  stellar populations revealed by early-type stars in LAMOST}


   \author{Chun Wang
          \inst{1}
          \and
          Haibo Yuan\inst{2}
          \and
          Maosheng Xiang\inst{3} \and Yuan-Sen Ting\inst{4,5} \and Yang Huang\inst{3} \and Xiaowei Liu\inst{6}
          }

   \institute{Tianjin Astrophysics Center, Tianjin Normal University, Tianjin 300387, People's Republic of China;
              \email{wchun@tjnu.edu.cn}
         \and
             Department of Astronomy, Beijing Normal University, Beijing 100875, People’s Republic of China;
             \email{yuanhb@bnu.edu.cn}
             \and 
             National Astronomical Observatories, Chinese Academy of Sciences, Beijing 100012, People’s Republic of China
             \and 
             Research School of Astronomy and Astrophysics, Australian National University, Cotter Rd., Weston, ACT 2611, Australia
             \and 
             School of Computing, Australian National University, Acton ACT 2601, Australia
             \and 
             South-Western Institute For Astronomy Research, Yunnan University, Kunming 650500, People’s Republic of China
             }

   \date{Received ---; accepted ---}

  \abstract{We investigate the radial metallicity gradients and azimuthal metallicity distributions on the  Galactocentric $X$--$Y$ plane using mono-temperature stellar populations selected from LAMOST MRS  young stellar sample. 
   The estimated radial metallicity gradient ranges from $-$0.015\,dex/kpc to $-$0.07\,dex/kpc, which decreases as effective temperature decreases (or stellar age increases)  at $7500 < T_{\rm eff} < 12500$\,K ($\tau < $1.5 Gyr). 
 The azimuthal metallicity excess (metallicity after subtracting radial metallicity gradient, $\Delta$\,[M/H]) distributions exhibit inhomogeneities with dispersions of 0.04\,dex to 0.07\,dex, which decrease as effective temperature decreases.   We also identify five potential metal-poor substructures with large metallicity excess dispersions. The metallicity excess distributions of these five metal-poor substructures suggest that they contain a larger fraction of metal-poor stars compared to other control samples. These metal-poor substructures may be associated with high-velocity clouds that infall into the Galactic disk from the Galactic halo, which are not quickly well-mixed with the pre-existing ISM of the Galactic disk. As a result, these high-velocity clouds produce some metal-poor stars and the observed metal-poor substructures.  
 The variations of metallicity inhomogeneities with different stellar populations indicate that high-velocity clouds are not well mixed with the pre-existing Galactic disk ISM within 0.3\,Gyr. }

\titlerunning{Metallicity variations}
\authorrunning{C. Wang}

\maketitle

%
\section{Introduction} \label{sec:intro}

Metals in the interstellar medium (ISM) regulate its' cooling process and the star formation of galaxies. The mixing process between infalling gas and the local ISM changes the fundamental properties, including the chemical abundance, of the local ISM and the future star formation process.  Understanding the mixing process of the ISM is crucial for informing our understanding of galaxy formation and evolution histories, such as Galactic archaeology and chemical evolution \citep{Tinsley1980, Matteucci2012, Matteucci2021}.  Most studies thus far assume that the ISM is well mixed at a small scale. 

The metallicity inhomogeneity of the ISM could help us to understand the ISM mixing.   At a large scale,  the ISM of the Galactic disk shows a negative radial metallicity gradient \citep{Balser}, suggesting an ``inside-out" disk formation scenario.  
 At a small scale,  the metallicity of the ISM \citep{Balser2011, Balser2015, DeCia2021}, Cepheid variable stars \citep{Pedicelli2009,  Poggio2022}, open clusters \citep{Davies2009,  Poggio2022},  young upper main-sequence stars \citep{  Poggio2022}, and young main-sequence stars \citep{Keith2022} in the  Milky Way (MW)  present azimuthal metallicity variations.   For other external spiral galaxies, people also found azimuthal ISM metallicity variations (after subtracting the radial metallicity gradients) by studying the metallicity of H\,II regions \citep{Ho2017,Ho2018,Kreckel2019,Kreckel2020}.  All these observed azimuthal metallicity inhomogeneities of the MW and other spiral galaxies suggest that the ISM  is not well mixed. To better understand the ISM mixing process,  we need to explore the variations of metallicity inhomogeneity of the ISM over time,  which has not been well studied yet. 

 Stellar surface chemical abundance remains almost unchanged during the main-sequence evolutionary stage.  It is thus a fossil record of the ISM  during the birth of these stars.  The metallicity inhomogeneity of  ISM and its variations over time can be studied using a stellar sample with a  wide age coverage. 
 However,  stars in the MW have moved away from their birth positions duo to the secular evolution of the MW.  Old stars have experienced stronger radial migration \citep{Minchev2010, Kubryk2015, Frankel2018, Frankel2020, Lian2022} compared to young stars.  According to \cite{Frankel2018} and \cite{Frankel2020},  68\% of stars have migrated within a distance of $3.6\sqrt{\tau/8 \rm\,{Gyr}}$\,kpc  and  $2.6\sqrt{\tau/6 \rm\,{Gyr}}$ kpc, respectively.   Hence, stars can move up $\sim$\,1\,kpc within $\,1$\,Gyr.  \cite{Lian2022} suggest that the average migration distance is 0.5--1.6\,kpc at
age of 2 Gyr and 1.0--1.8 kpc at age of 3 Gyr. 
 In conclusion, young stars in the MW are valuable for studying the ISM mixing process due to their large sample size and wide age coverage compared to other young objects (e.g., open clusters, Cepheid variable stars, H\,II regions)and because they are less affected by secular evolution compared to older stars.   For young main-sequence stars,  effective temperatures are tightly correlated with stellar ages  \citep{Schaller1992, Zorec2012, Sun2021}.  The variations of ISM metallicity homogeneity (or inhomogeneity)  over time can be investigated using a young main-sequence stellar sample with accurate determinations of stellar atmospheric parameters  (effective temperature $T_{\mathrm{eff}}$, surface gravity $\log g$ and metallicity [M/H]). 


Recently, accurate stellar atmospheric parameters were derived from  LAMOST medium-resolution spectra for  40,034 young (early-type) main-sequence stars \citep{Sun2021}.  The stellar sample spans a wide range of effective temperatures (7500-15000K), and covers a large and contiguous volume of the Galactic disk.  It allows us to investigate the ISM metallicity inhomogeneity and its variations over time. 
In this work, we investigate the radial metallicity gradients and azimuthal metallicity distribution features of mono-temperature stellar populations across the Galactic disk within $-10.5 < X < -8.0$\,kpc, $-1.0 < Y < 1.5$\,kpc and $|Z| < 0.15$\,kpc using the young stellar sample. 

This paper is organized as follows. In Section\,2, we introduce the adopted young stellar sample. In Section\,3, we present the radial metallicity gradients of mono-temperature stellar populations.  We present the azimuthal metallicity distributions of mono-temperature stellar populations in Section\,4.  The constraints on the ISM mixing process using the azimuthal metallicity distributions are discussed in Section\,5. Finally, we summarize our work in Section\,6. 



\section{ Young stellar sample from the LAMOST Medium-Resolution Spectroscopic Survey }

\subsection{Coordinate Systems and Galactic Parameters}
 Two coordinate systems are used in this paper.  One is a right-handed Cartesian coordinate system ($X, Y, Z$) centred on the Galactic centre, with $X$ increasing towards the Galactic centre, $Y$ in the direction of Galactic rotation, and $Z$ representing the height from the disk mid-plane, positive towards the north Galactic pole. The other is a Galactocentric cylindrical coordinate system $(R, \Phi, Z)$, with $R$ representing the Galactocentric distance,
$\Phi$ increasing in the direction of Galactic rotation,  and $Z$ the same as that in the Cartesian system.
The Sun is assumed to be at the
Galactic midplane (i.e., $Z_{\odot} =$ 0\,pc) and has a value of $R_{\odot}$ equal to
8.34\,kpc \citep{Reid2014}. 


\subsection{Sample selections}

 As of March 2021, the LAMOST Medium-Resolution Spectroscopic Survey had collected 22,356,885 optical (4950--5350\,\AA\,and 6300--6800\,\AA) spectra with a resolution of $R\sim 7500$.  
\cite{Sun2021}  selected 40,034 late-B and A-type main-sequence stars from the LAMOST Medium-Resolution Spectroscopic Survey (hereafter named the LAMOST-MRS young stellar sample) and extracted their accurate stellar atmospheric parameters.  
For a star with spectral SNR\,$\sim$60, the cross validated scatter is $\sim$ 75\,K, 0.06\,dex and 0.05\,dex for $T_{\mathrm{eff}}$, $\log g$ and  [M/H], respectively.   We adopt the photogeometric distances provided by \cite{gaiadistance} for these stars.   The $X, Y, Z, R, \Phi$ of each star are estimated using its distance and coordinates, and the corresponding errors are also given using the error transfer function. 

\begin{figure}
\centering
\includegraphics[width=3.5in]{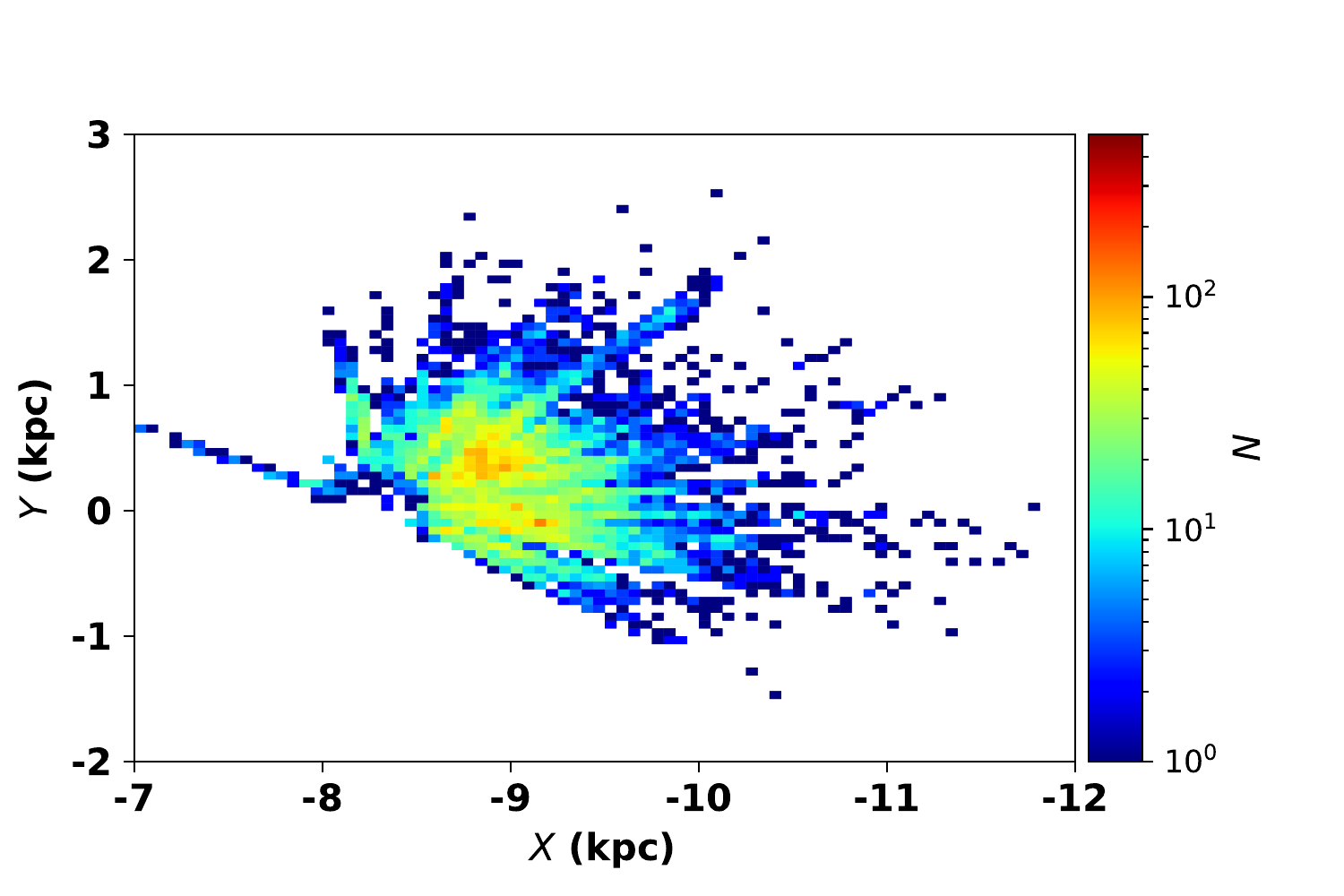}
\caption{ Stellar number density distribution on the $X$--$Y$ plane of the  final LAMOST-MRS young stellar sample. }
\label{sample_distributions}
\end{figure}

For the LAMOST-MRS young stellar sample, stars with $T_{\mathrm{eff}}$ and [M/H] uncertainties larger than 75\,K and 0.05\,dex are discarded to ensure the reliability of the results.  Only stars with $7000 < T_{\mathrm{eff}}  < 15000$\,K,  $\log g > 3.5$\,dex,  and $-1.2 < $ [M/H] $ < 0.5$\,dex are selected, as they are main-sequence stars with accurate metallicity determinations.
 Because young stars are mostly located in the Galactic plane,  stars of  $|Z| \geq 0.15$\,kpc are also removed to reduce the contaminations from old stars.  Finally, our LAMOST-MRS young stellar sample consists of 14,692 unique stars. The stellar number density distribution on the $X$--$Y$ plane is shown in Figure\,\ref{sample_distributions}. 
  
 The faint limiting magnitude of the LAMOST-MRS is $g \sim$\,15\,mag. The faintest stars in the four effective temperature bins (adopted in the following context) have absolute $g$-band  magnitudes of $\sim$\,4.5\,mag ($7500 < T_{\mathrm{eff}} < 8000$\,K), $\sim$\,4.0\,mag ($8000 < T_{\mathrm{eff}} < 8500$\,K), $\sim$\,3.5\,mag ($9500 < T_{\mathrm{eff}} < 10500$\,K),  and $\sim$\,3.2\,mag ($10000 < T_{\mathrm{eff}} < 12500$\,K).  Based on the distance modulus,  stellar samples with $7500 < T_{\mathrm{eff}} < 8000$\,K, $8000 < T_{\mathrm{eff}} < 8500$\,K, $9500 < T_{\mathrm{eff}} < 10500$\,K and $10000 < T_{\mathrm{eff}} < 12500$\,K are incomplete at distance ($d$)$\,>$\,1.258\,kpc, $d >$\,1.584\,kpc, $d >$\,1.955\,kpc and $d >$\,2.29\,kpc, respectively. 
 

 \subsection{Correcting for the Teff-metallicity systematics caused by NLTE effect} 

\begin{figure*}
\centering
\includegraphics[width=6.5in]{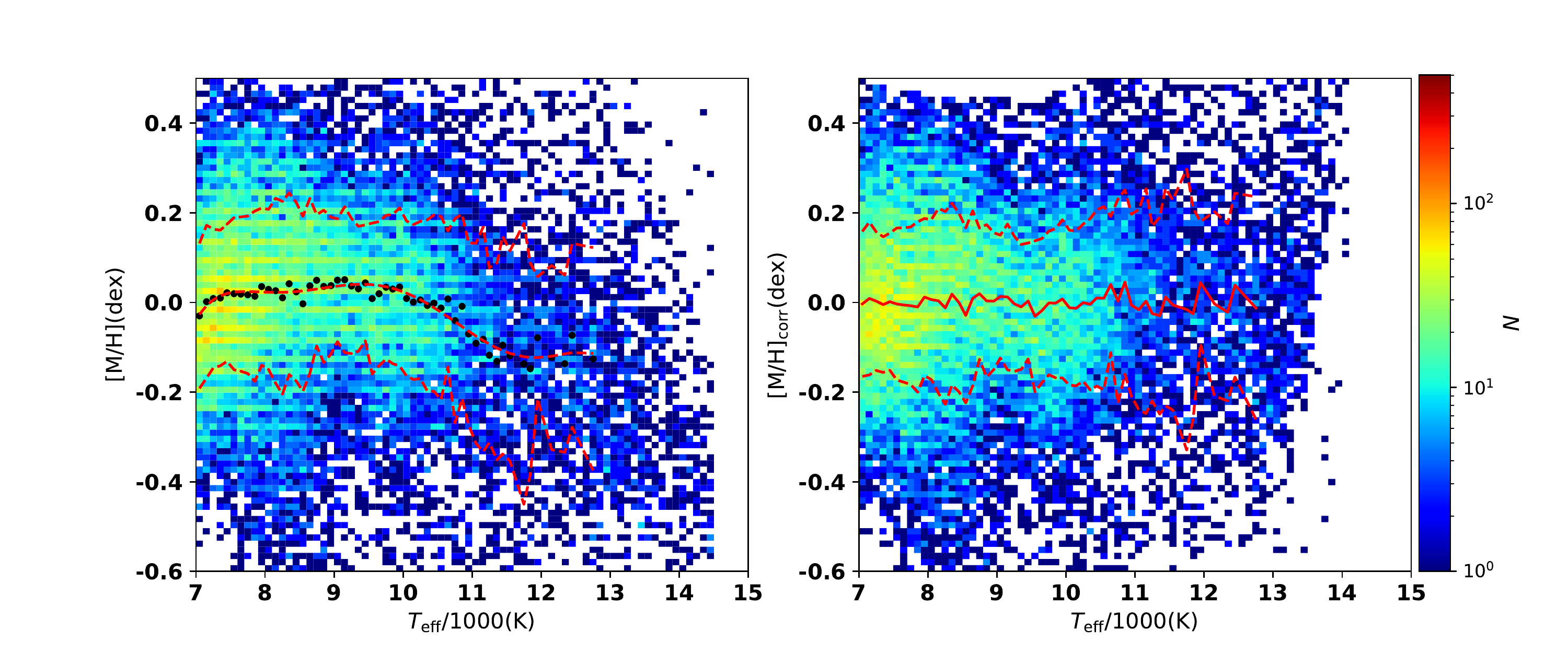}
\caption{ The stellar number density distributions on the effective temperature and metallicity planes. The left panel displays the relation between the estimated metallicity and $T_{\mathrm{eff}}$ (black symbols) and their corresponding polynomial fit  (red solid line) for the LAMOST-MRS young stellar sample.   The black symbol represents the mean estimated metallicity and $T_{\mathrm{eff}}$ in each $T_{\mathrm{eff}}$ bin with a $\Delta_{T_{\mathrm{eff}}} \rm{of}\,100$\,K.
The dashed red lines represent the corresponding standard deviations of the metallicity in each $T_{\mathrm{eff}}$ bin.   Additionally, the stellar number density distribution on the $T_{\mathrm{eff}}$-$\rm [M/H]$ plane is plotted in the left panel. Right panel shows the stellar number density distribution on the $T_{\mathrm{eff}}$-$\rm [M/H]_{corr}$ plane.  The solid and dashed red lines represent the mean corrected metallicity values and the corresponding standard deviations of the metallicity in each $T_{\mathrm{eff}}$ bin, respectively.  }
\label{feh_teff}
\end{figure*}

Metallicities of  LAMOST-MRS young stars are estimated based on LTE models.  But the NLTE effects can significantly affect their spectra.  The NLTE effects may incur a bias of estimated metallicity with respect to $T_{\mathrm{eff}}$ \citep{Xiang2021}. To reduce the NLTE effects on the estimated metallicity values,  we use a sixth-order polynomial to model the $T_{\mathrm{eff}}$--metallicity trend of the LAMOST-MRS young stellar sample.  The dependency of metallicity on $T_{\mathrm{eff}}$ is mitigated by subtracting the fitted relation.  We fit the relation of metallicity and $T_{\mathrm{eff}}$  only using stars of $8.7 < R$ $< 9.5$\,kpc to avoid the effects of the radial metallicity gradients.   Figure\,\ref{feh_teff} shows the polynomial fit of estimated $T_{\mathrm{eff}}$--metallicity trend. 
The distribution of stars on the $T_{\mathrm{eff}}$--$\rm [M/H]_{corr}$ plane is also presented in Figure\,\ref{feh_teff}, where the $\rm [M/H]_{corr}$ is the metallicity after subtracting the trend. 
 The figure shows that the final metallicity ($\rm [M/H]_{corr}$) does not vary with $T_{\mathrm{eff}}$ after subtracting the mean trend.   The used metallicity is the corrected metallicity ($\rm [M/H]_{corr}$) in the following context. 

\section{Radial metallicity gradients of mono-temperature young stellar populations}
\begin{figure*}
\centering
\includegraphics[width=6.5in]{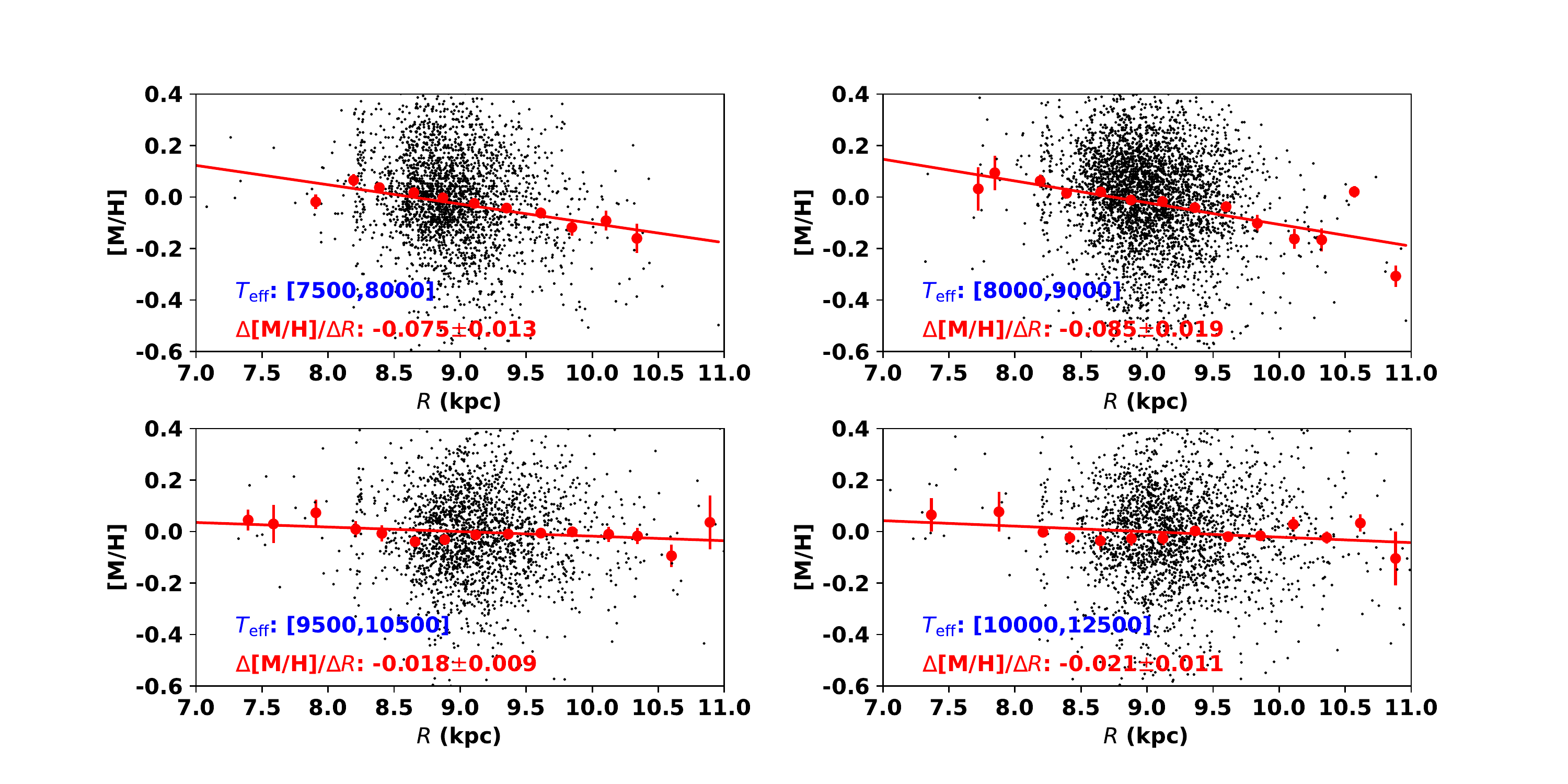}
\caption{Radial metallicity gradients of these four mono-temperature stellar populations. The red symbols represent the median metallicity values in individual radial bins.  The line in red represents the linear regression over the red symbols. The temperature range of each stellar population, the slope of the linear fit (the
radial metallicity gradient), and its associated uncertainty are marked at the bottom left panel.}
\label{metallicity_gradients_fit}
\end{figure*}

In this section, we investigate the radial metallicity gradients of mono-temperature stellar populations.  We divide stars from the  LAMOST-MRS stellar sample into different stellar populations in four effective temperature bins: $7500 < T_{\mathrm{eff}} < 8000$\,K,  $8500 < T_{\mathrm{eff}} < 9000$\,K,  $9500 < T_{\mathrm{eff}} < 10500$\,K,  and  $10000 < T_{\mathrm{eff}} < 12500$\,K. We discard stars with   $9000 < T_{\mathrm{eff}} < 9500$\,K because the  $\rm H{\alpha}$ lines of these stars are not sensitive with $T_{\mathrm{eff}} $, which leads to large uncertainties of the measured metallicities.   In each temperature bin, we further divide stars into a small radial annulus of 0.25 kpc. Bins containing fewer than 3 stars are discarded.  We perform linear regression on the metallicity as a function of the Galactic radius $R$. The slope is adopted as the radial metallicity gradient.  Figure\,\ref{metallicity_gradients_fit}  shows the fitting results for these four mono-temperature stellar populations.  As shown in the figure, a linear regression captures the trend between metallicity and Galactic radius $R$ well. 

The radial metallicity gradient of the Galactic disk plays an important role in the study of Galactic chemical and dynamical evolution histories. It is also a fundamental input parameter in any models of Galactic chemical evolution.  Previous studies on the radial metallicity gradient of the Galactic disk using different tracers (including   OB stars by   \cite{daflon}, Cepheid variables by \cite{Andrievsky, Luck}, H~{\sc ii} regions by \cite{Balser},  open clusters by \cite{Chen,Magrini}, planetary nebulae by \cite{Costa,Henry} , FGK dwarfs by \cite{Katz,Cheng,Boeche}, red giant stars by \cite{Hayden_2014,Boeche2014} and red clump stars by \cite{Huang}) have presented that the Galactic disk has a negative radial metallicity gradient, which generally supports an    ``inside-out" disk-forming scenario. The variations of radial metallicity gradient with stellar age ($\tau$) have also been investigated by \cite{Xiang2015} and \cite{Wang2019a} using main-sequence turn-off (MSTO) stars and other works \citep[e.g.,][]{Casagrande2011, Toyouchi2018}.   \cite{Wang2019a} found that  radial metallicity gradients steepen with increasing age at $\tau <$\,4\,Gyr, reach a maximum at $4 < \tau < 6$ Gyr , and then flatten with age.  \cite{Xiang2015} also found a similar trend of radial metallicity gradients with stellar ages. The relation of radial metallicity gradient with stellar age for the youngest stellar populations ($\tau < 2$ Gyr) was not investigated in these two works due to the inaccuracy of estimated stellar ages for the youngest MSTO stars.

\begin{figure}
\centering
\includegraphics[width=3.5in]{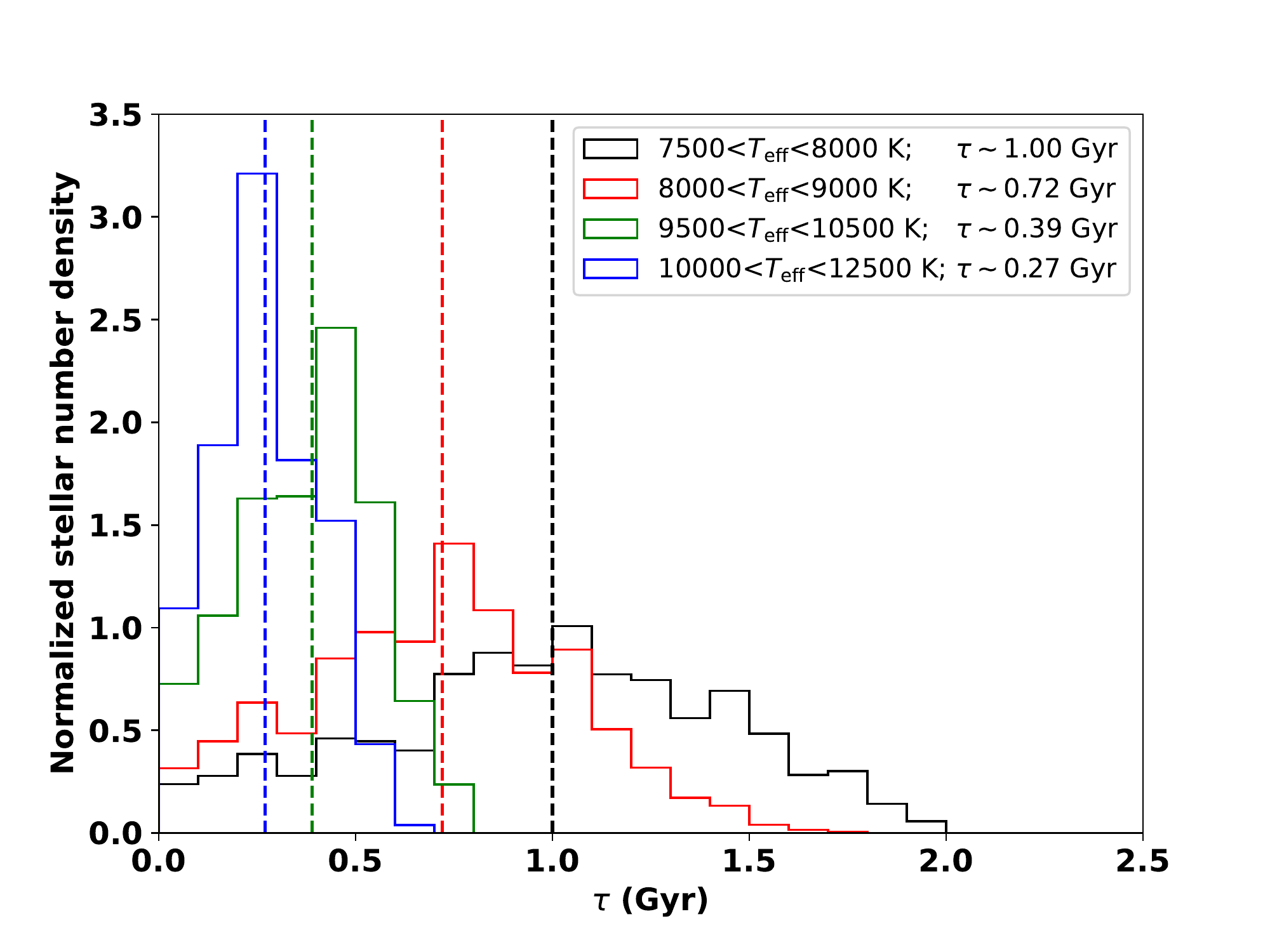}
\caption{The age distributions of stars within the  metallicity range of $-0.5$\,dex and 0.5\,dex   in these four effective temperature bins of  $7500 < T_{\mathrm{eff}} < 8000$\,K,  $8500 < T_{\mathrm{eff}} < 9000$\,K,  $9500 < T_{\mathrm{eff}} < 10500$\,K, and  $10000 < T_{\mathrm{eff}} < 12500$\,K predicted by the  PARSEC isochrones.  The median ages are over-plotted with dashed lines. The effective temperature ranges and median ages for each bin are marked at the top right corner of the figure. }
\label{age_teff}
\end{figure}

For the youngest main-sequence stars, the stellar age has a tight relation with the effective temperature.  We derive median stellar ages for these four mono-temperature stellar populations according to the  PARSEC isochrones \citep{Bressan2012}. 
Age distributions in these four effective temperature bins with the metallicity range of $-0.5$\,dex and 0.5\,dex (similar to the metallicity coverage of our sample) predicted by the  PARSEC isochrones are shown in Figure\,\ref{age_teff}.  The median ages are 1.00 Gyr, 0.72 Gyr, 0.39 Gyr, and 0.27 Gyr, respectively, in these four effective temperature bins of $7500 < T_{\mathrm{eff}} < 8000$\,K,  $8500 < T_{\mathrm{eff}} < 9000$\,K,  $9500 < T_{\mathrm{eff}} < 10500$\,K, and  $10000 < T_{\mathrm{eff}} < 12500$\,K.  

\begin{figure}
\centering
\includegraphics[width=3.5in]{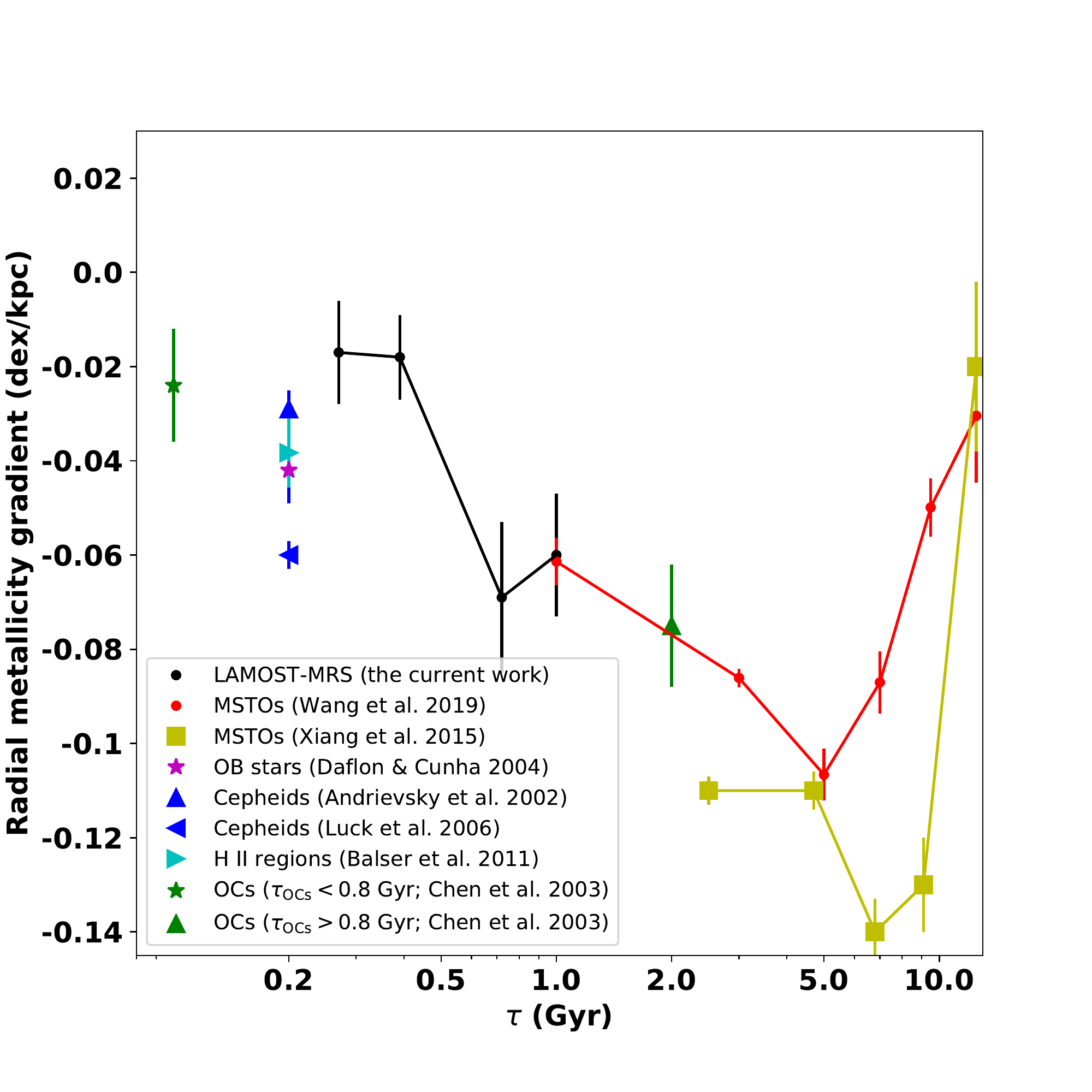}
\caption{ Radial metallicity ([M/H]/[Fe/H]) gradient variations' with stellar ages of the  LAMOST-MRS  young stellar sample in this work and these of MSTO presented by  \cite{Wang2019a} and \cite{Xiang2015}. Radial metallicity gradients estimated using young objects (OB stars, Cepheid variables, H~{\sc ii} regions and OCs) from previous works are also shown using different symbols as labelled in the figure.  Note that the $X$-axis is logarithmic.}
\label{compare_metallicity_gradients}
\end{figure}

 Figure\,\ref{compare_metallicity_gradients} shows the variations of radial metallicity gradients with the stellar ages of  LAMOST-MRS young stellar sample by us, the stellar ages of MSTO  stars presented by \cite{Wang2019a} and \cite{Xiang2015}.  Radial metallicity gradients estimated using young objects (including OB stars, Cepheid variables, H~{\sc ii} regions and open clusters (OCs)) are also over-plotted in the figure.  We assume a mean age of 0.2 Gyr for OB stars, Cepheid variables, H~{\sc ii} regions.  \cite{Chen} divided their OCs into two age bins: $\tau < 0.8$\,Gyr and $\tau > 0.8$\,Gyr.  We assume a mean age of 0.1 Gyr and 2 Gyr following \cite{Chen}.
  Radial metallicity gradients estimated using our LAMOST-MRS young stellar sample are well aligned with those values estimated using other young objects of previous works \citep{daflon, Andrievsky, Luck, Balser, Chen} within the uncertainties of the measurements. 
 We find that radial metallicity gradients of young stellar populations ($\tau < 2$\,Gyr) steepen as age increases  (with a gradient of $\sim -0.055\,\rm{dex\,kpc^{-1}\,Gyr^{-1}}$) within the uncertainty of our estimates.  The result is consistent with those of \cite{Wang2019a} and \cite{Xiang2015}, who also found that the radial metallicity gradients decrease with increasing stellar age for young stellar populations (1.5 $< \tau <$ 4 Gyr). 
 Our results extend the relation between radial metallicity gradient and stellar age to $\tau <$\,1.5\,Gyr compared to \cite{Wang2019a} and \cite{Xiang2015}.
 
\section{Azimuthal metallicity distributions of mono-temperature stellar populations}

 This section aims to investigate the azimuthal metallicity distributions of different mono-temperature stellar populations, as divided in Section\,3. Doing so, we further divide all these four mono-temperature stellar populations into different $X$--$Y$ bins with a bin size of 0.1$\times$0.1\,kpc.  This chosen spatial bin size aims to strike a balance between spatial resolution and the number of stars in each bin, with a preference for higher spatial resolution and a larger number of stars in each bin.  Figure\,\ref{number_xy_all_bootstrap} displays the distributions of stellar numbers in all $X$--$Y$ bins. The typical uncertainties of $X$ and $Y$ are respectively 0.0114\,kpc and 0.0055\,kpc, which are much smaller than the bin size of 0.1\,kpc.  To better visualize the azimuthal variation, we subtract the mean radial metallicity gradient as measured in the previous section. 
 We introduce a new metallicity scale denoted as the "Metallicity Excess",  $\Delta \rm [M/H] = [M/H] - [M/H]_{R}$,  where $\rm [M/H]_{R}$ is the median metallicity in each $R$ bin with a bin scale of 0.5\,kpc.  In each $X$ and $Y$ bin,  we estimated the mean $\Delta \rm [M/H]$  and its associated dispersion ($\sigma \rm [M/H]$), as well as their uncertainties.

\subsection{Global statistics}

\begin{figure*}
\centering
\includegraphics[width=6.5in]{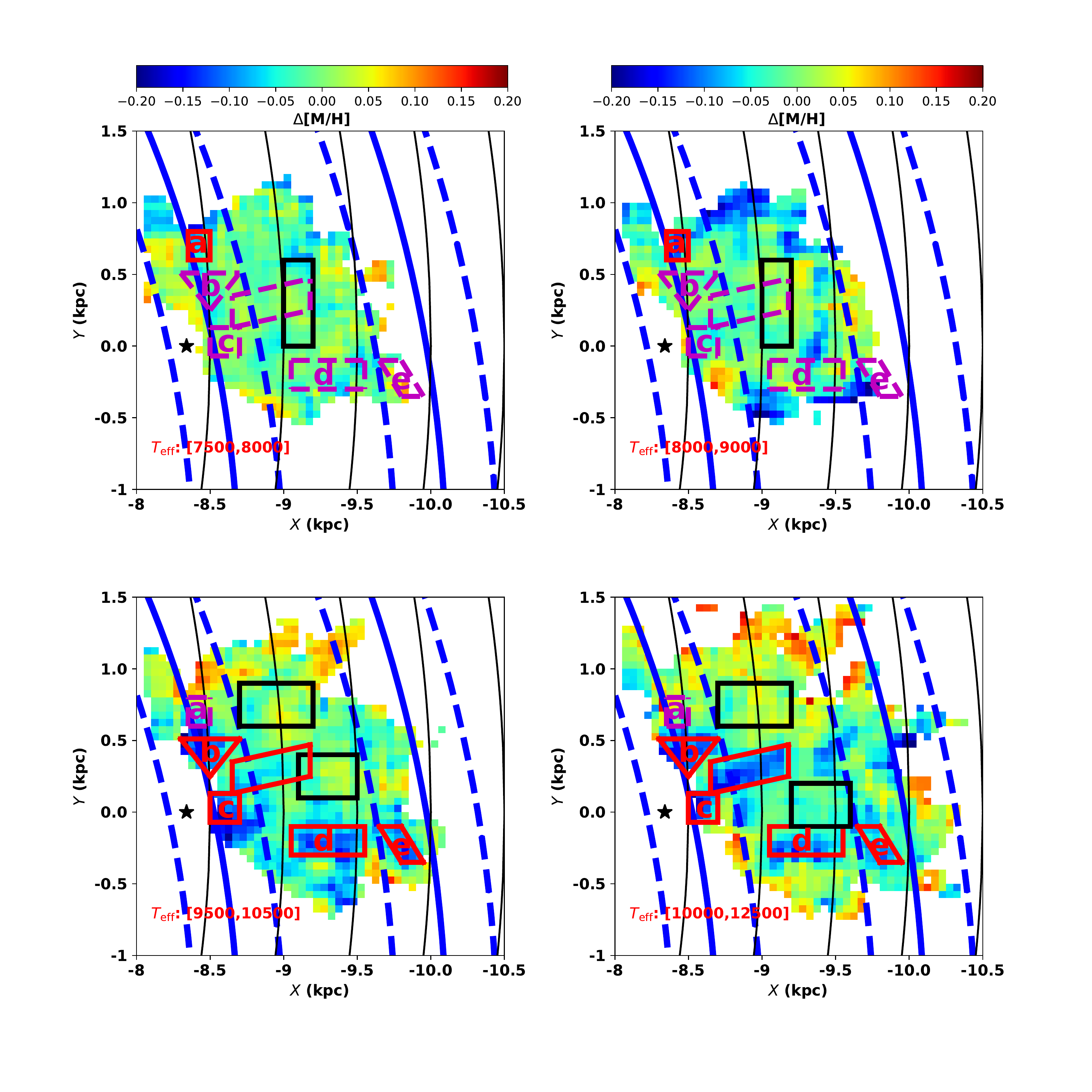}
\caption{Metallicity excess distributions after subtracting radial metallicity gradients for these four mono-temperature  stellar populations, binned
by 0.1$\times$0.1 kpc on the $X$--$Y$ plane.  Bins that contained fewer than eight stars are discarded.  The minimum number of stars in each bin is selected to ensure that the median metallicity excess uncertainties produced in these bins could distinguish the metal-poor substructures from other regions. The positions at $R=$\,8.5, 9.0, 9.5, 10.0, and 10.5\,kpc are marked with black arcs. The black star symbol indicates the position of the Sun.  The centre and 1$\sigma$ width of the spiral arms are shown by blue solid and dashed lines in all panels. Regions framed in red lines and magenta dashed lines indicate the spatial positions of these five metal-poor substructures (labelled by `a', `b', `c', `d', and `e').   A region framed by red lines means that it is a real metal-poor substructure of this mono-temperature stellar population. The region framed by magenta dashed lines is a corresponding region of metal-poor substructure in other mono-temperature stellar populations. Control regions are framed in black lines. }
\label{metallicity_xy_all_bootstrap}
\end{figure*}

\begin{figure}
\centering
\includegraphics[width=3.5in]{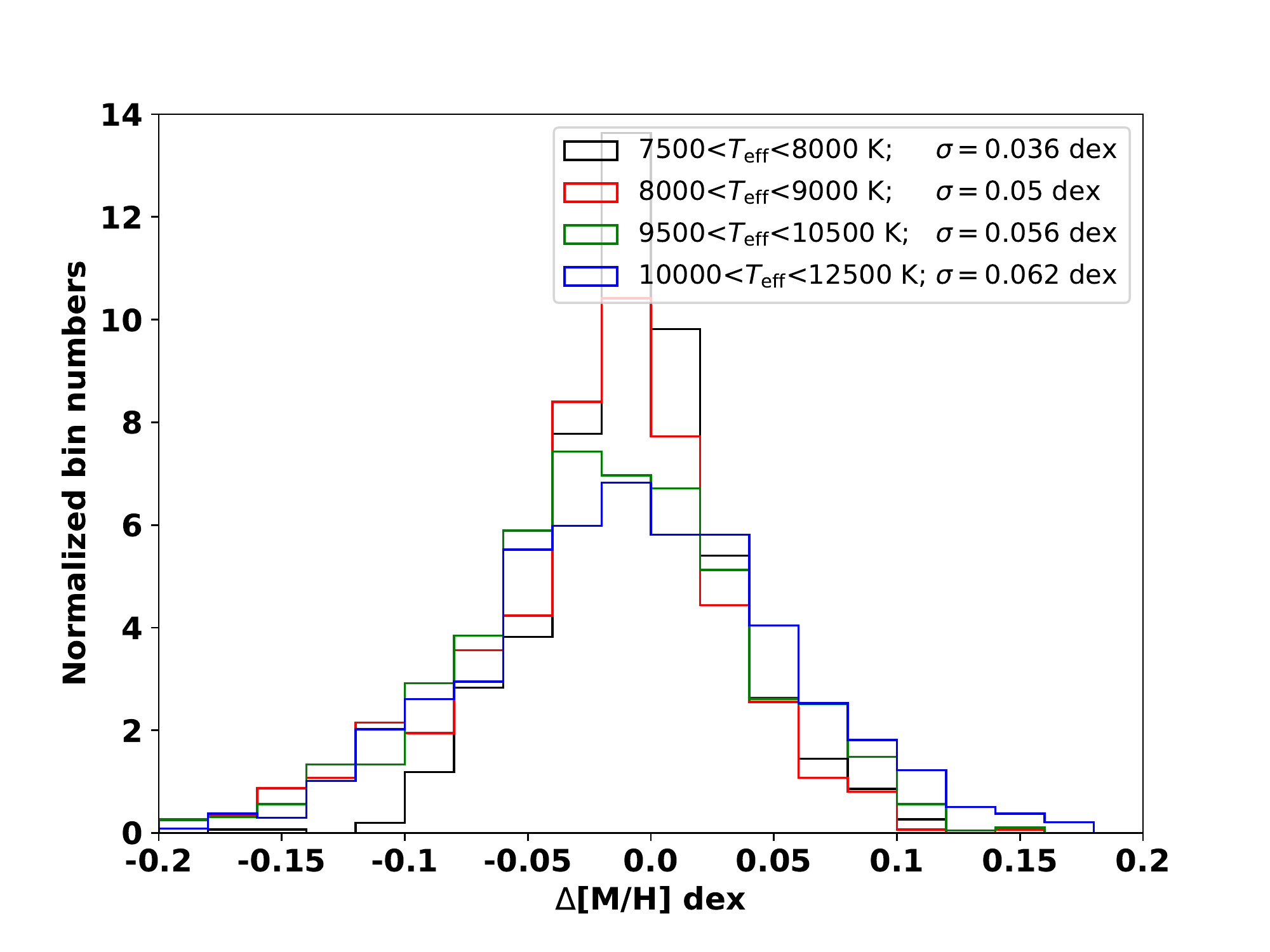}
\caption{Bin number density distributions of metallicity excess in Figure\,\ref{metallicity_xy_all_bootstrap} of these four mono-temperature  stellar populations. The effective temperature ranges and dispersions of these four distributions are marked at the right top corner of the figure.}
\label{metallicity_dispersion}
\end{figure}

Figure\,\ref{metallicity_xy_all_bootstrap}  shows the azimuthal metallicity excess ($\Delta \rm [M/H]$) distributions of these four mono-temperature stellar populations on the $X$--$Y$ plane, which shows significant inhomogeneities on the bin scale of 0.1$\times$0.1\,kpc.   The difference of metallicity excess spans almost 0.4\,dex (ranging from $-0.2$\,dex to 0.2\,dex).   The dispersions of the (median) metallicity excess (shown in Figure\,\ref{metallicity_xy_all_bootstrap}), are respectively 0.04\,dex, 0.058\,dex, 0.057\,dex, and 0.066\,dex for the stellar populations with effective temperature coverage of   $7500 < T_{\mathrm{eff}} < 8000$\,K,  $8500 < T_{\mathrm{eff}} < 9000$\,K,  $9500 < T_{\mathrm{eff}} < 10500$\,K, and  $10000 < T_{\mathrm{eff}} < 12500$\,K as shown in Figure\,\ref{metallicity_dispersion}.  The metallicity excess dispersion increases as temperature increases. 
 
We compute the two-point correlation function of metallicity excess for these four mono-temperature stellar populations to quantify the scale length associated with the observed
metallicity inhomogeneity with the same method of \cite{Kreckel2020}. The 50\% level scale length is $\sim$ 0.05\,kpc, which is smaller than our spatial bin size of 0.1\,kpc. The result suggests that the scale length of metallicity inhomogeneity is smaller than 0.1\,kpc.  This scale length is much smaller than the  0.5-1.0kpc predicted by \cite{Krumholz2018} and the 0.3\,kpc (50\% level) estimated by \cite{Kreckel2020} using HII regions in external galaxies.  It is possible that their results in external galaxies may not be directly applicable to the Milky Way.  The origin of the chemical inhomogeneities observed in this paper may differ from those observed in external galaxies.   The scale length is consistent with that of \cite{DeCia2021}, who suggests that pristine gas falling into the "Galactic disk" can produce chemical inhomogeneities on the scale of tens of pc.  
 
 To check the reliability of metallicity inhomogeneities, we study the distributions of metallicity excess uncertainties ($\Delta \rm [M/H]_{err}$)  in Figure\,\ref{metallicity_xy_all_bootstrap_err}.   $\Delta \rm [M/H]_{err}$ are mostly smaller than 0.1\,dex (with median values of  0.030, 0.039, 0.042, and 0.048 dex for stellar populations of $7500 < T_{\mathrm{eff}} < 8000$\,K,  $8500 < T_{\mathrm{eff}} < 9000$\,K,  $9500 < T_{\mathrm{eff}} < 10500$\,K, and  $10000 < T_{\mathrm{eff}} < 12500$\,K, respectively), which is much smaller than the difference of metallicity excess shown in Figure\,\ref{metallicity_xy_all_bootstrap}. The comparison between  Figure\,\ref{metallicity_xy_all_bootstrap} and Figure\,\ref{metallicity_xy_all_bootstrap_err} suggests that the observed azimuthal metallicity inhomogeneities are reliable. 
 
 Figure\,\ref{metallicity_xy_all_bootstrap} and Figure\,\ref{metallicity_dispersion} not only show the azimuthal metallicity inhomogeneity but also its variations with effective temperature.  Azimuthal metallicity inhomogeneities (larger metallicity dispersion means larger metallicity inhomogeneity) of young stellar populations ($T_{\mathrm{eff}} > 8000$\,K) are much larger than that of the old stellar population ($T_{\mathrm{eff}} < 8000$\,K).

Besides metallicity excess distributions, we also investigate the spatial distributions of metallicity excess dispersions. Figures\,\ref{sigma_metallicity_xy_all} and \ref{sigma_metallicity_xy_all_err} show the spatial distributions of metallicity excess dispersion and its associated uncertainty.  From Figure\,\ref{sigma_metallicity_xy_all}, we can find inhomogeneities of metallicity dispersion distributions for the four mono-temperature stellar populations. Inhomogeneities of young stellar populations ($T_{\mathrm{eff}} > 8000$\,K)  are much larger than those of the old stellar population ($T_{\mathrm{eff}} < 8000$\,K).  
Metallicity dispersion uncertainties shown in Figure\,\ref{sigma_metallicity_xy_all_err} are much smaller than the difference of metallicity dispersions shown in Figure\,\ref{sigma_metallicity_xy_all}, which suggests that observed azimuthal metallicity dispersion distributions in Figure\,\ref{sigma_metallicity_xy_all} are reliable.

\subsection{Metal-poor substructures}
From Figure\,\ref{metallicity_xy_all_bootstrap}  and  Figure\,\ref{sigma_metallicity_xy_all}, we observe several metal-poor substructures with large metallicity dispersions. 
In the old stellar population ($8000<T_{\mathrm{eff}} < 9000$\,K), we identify one metal-poor substructure (labelled `a' in  Figure\,\ref{metallicity_xy_all_bootstrap}), which is not found in the stellar populations of $T_{\mathrm{eff}} > 9500$\,K.  The metallicity of this metal-poor substructure is smaller than its nearby regions by $\sim$\,0.1\,dex.  The metal-poor substructure is also found in the stellar population of $7500<T_{\mathrm{eff}} < 8000$\,K, while it is much weaker.   In the younger stellar population ($9500 < T_{\mathrm{eff}} < 12500$\,K),  there are four metal-poor regions labelled  `b', `c', `d, and `e'   in Figure\,\ref{metallicity_xy_all_bootstrap}, which are not found in the two old stellar populations. The metallicities of these four regions are also smaller than their nearby regions by $\sim$\,0.1\,dex.    The spatial sizes of `a', `b', `c', `d',  and `e' substructures are respectively $\sim 0.7\times0.2$, $0.3\times0.2$, $0.8\times0.2$, $0.4\times0.1\,\rm kpc$, and $0.1\times0.4\,\rm kpc$.    

Azimuthal metallicity variations with spatial positions may be tracked along spiral arms \citep[e.g.,][]{Khoperskov2018, Poggio2022}. We compare the azimuthal metallicity distributions, especially those of the five identified metal-poor substructures, with the spiral arms (shown as the blue solid and dashed line, indicating from left to right segments of the Local (Orion) arm and the Perseus arm)  determined using high mass star forming regions by \cite{Reid2019}.  From  Figure\,\ref{metallicity_xy_all_bootstrap}, we find that metallicity distributions of all these four mono-temperature stellar populations do not track the expected locations of spiral arms, similar to the results presented by \cite{Keith2022}.  \cite{Keith2022} mapped out azimuthal metallicity distributions using LAMOST OBAF young stars selected from LAMOST low-resolution spectra by \cite{Xiang2021}.  However, our detailed azimuthal metallicity patterns have large differences compared to those of \cite{Keith2022}.  We cross-match the LAMOST-MRS young stellar sample with their OBAF stellar sample and plot azimuthal metallicity distributions using the common stars. For these common stars, the azimuthal metallicity distributions are similar either using the metallicity of the LAMOST-MRS or LAMOST OBAF sample, which are also similar to the results presented here.   Different fractions and the origin of contaminations from cold stars of the two samples may be responsible for the difference between the results of us and those of \cite{Keith2022}.

In Figure\,\ref{metallicity_dispersion_compare} we show the normalised metallicity excess distributions of these five metal-poor regions (`a', `b', `c, `d',  and `e' regions) and those of control regions (framed in black lines).  The skewnesses of the metallicity distributions are also estimated.  All these skewness are smaller than zero, suggesting that metallicity distributions, both in the metal-poor regions and control regions, have metal-poor tails. Metal-poor substructures contain a larger fraction of metal-poor stars and have much smaller skewness compared to control regions.


\section{The ISM mixing process}
Azimuthal metallicity distributions mapped out by the LAMOST-MRS  stellar sample presented in Section\,4  suggest that there are significant azimuthal metallicity inhomogeneities.  Similarly,  the azimuthal metallicity distributions of the ISM \citep{Balser2011, Balser2015, DeCia2021}, Cepheid variable stars \citep{Pedicelli2009}, and open clusters\citep{Davies2009, Fu2022}  in the MW also show significant azimuthal metallicity inhomogeneities.  
 Azimuthal metallicity inhomogeneities of external spiral galaxies are also found through investigating the metallicity of H\,II regions \citep{Ho2017,Ho2018,Kreckel2019,Kreckel2020}.  These observed azimuthal metallicity inhomogeneities both in the MW and external spiral galaxies suggest that the ISM is not well mixed.

 High-velocity clouds infalling into the Galactic disk from the Galaxy halo are metal-poor, with metallicities ranging from 0.1 to 1.0 solar metal \citep{Foxetal.2017,Wright2021,DeCia2021}.  High-velocity clouds can be generated by the ``galactic fountain" cycle and other gas accretion processes.  Intermediate-velocity clouds are also the main observational manifestations of the ongoing ``galactic fountain" cycle \citep{Wakker1997}. However, unlike high-velocity clouds, intermediate-velocity clouds are nearby systems and have near or solar metallicity. 
These five metal-poor substructures found in the current paper have smaller mean metallicity values and larger dispersions. They also contain a larger fraction of metal-poor stars than other Galactic disk regions. These results suggest that these five metal-poor substructures may be associated with high-velocity clouds, which infall into the Galactic disk from the Galactic halo.  These high-velocity clouds are not quickly well mixed with the ISM of the Galactic disk after they infall into the Galactic disk,  thus they have time to produce stars, which are more metal-poor than stars born in the pre-existing Galactic disk ISM. As a result, the metallicity dispersions in the locations of these high-velocity clouds are larger than those of other regions.
 


As shown in Section\,4, the azimuthal metallicity inhomogeneity and dispersion increase as temperature increases.   
Five metal-poor (large metallicity dispersion) substructures are found. Four of these substructures are only found in the metallicity distributions of young stellar populations ($T_{\rm eff} > 9500\, \rm K$).   
These metal-poor substructures of the stellar population with $9500 < T_{\rm eff} < 10500\, \rm K$ (with a median stellar age of 0.39 Gyr) may suggest that the corresponding high-velocity clouds had infalled into the Galactic disk  0.39 Gyr ago. 
Similarly, these metal-poor substructures of  $10000 < T_{\rm eff} < 12500\, \rm K$ (with a median stellar age of 0.27 Gyr) may suggest that these corresponding high-velocity clouds had not been well mixed into the Galactic disk ISM  before 0.27\,Gyr. 
 In conclusion, these high-velocity clouds infalling into the Galactic disk from the Galactic halo are not well mixed with the pre-existing Galactic disk ISM within 0.12\,Gyr ($0.39-0.27$\,Gyr).

It is noted that the  `d' metal-poor substructure of the stellar population with $10000 < T_{\rm eff} < 12500\, \rm K$ move $\sim 0.1$\,kpc on the $X$--$Y$ plane compared to that of  $9500 < T_{\rm eff} < 10500\, \rm K$.  This may be the consequence of the velocity difference between infalling high-velocity clouds and stars or ISM of the Galactic disk. A velocity difference of $\sim 0.82\,\rm km s^{-1}$ is needed to move 0.1\,kpc within 0.12 Gyr, which is reasonable. 

The other metal-poor substructure is found in the metallicity distributions of old stellar populations ($T_{\rm eff} < 9000\, \rm K$), suggesting that the corresponding high-velocity cloud had not sufficiently mixed with the Galactic disk ISM  before 0.72\,Gyr (the median age of the stellar population with $8000 < T_{\rm eff} < 9000\, \rm K$) and had infalled into the Galactic disk 1.0\,Gyr (the median age of the stellar population with $7500 < T_{\rm eff} <8000 \, \rm K$) ago.  The results suggest that the high-velocity cloud is not well mixed into Galactic disk ISM within 0.28\,Gyr. This substructure is not found in the metallicity distributions of young stellar populations, suggesting that this high-velocity cloud had been well mixed into the Galactic disk ISM  0.39\,Gyr ago. 

\section{Summary}
In this work, we use the LAMOST MRS young stellar sample with accurate effective temperature and metallicity to investigate the radial metallicity gradients and azimuthal metallicity distributions of different stellar populations with varying effective temperatures (or ages).

The estimated radial metallicity gradient ranges from $-$0.015\,dex/kpc to $-$0.07\,dex/kpc, which decreases as effective temperature decreases (or stellar age increases).  The result is consistent with those of \cite{Xiang2015} and \cite{Wang2019a}, who also found that the radial metallicity gradients decrease with increasing stellar age for young stellar populations ($1.5 <\,\tau < 4\,\rm Gyr$).  Our results extended their study of the relation between radial metallicity gradient and stellar age to $\tau < 1.5\,\rm Gyr$.

After subtracting radial metallicity gradients on the $X$--$Y$ plane,  the azimuthal metallicity distributions of all these four mono-temperature stellar populations show significant metallicity inhomogeneities,  which is consistent with previous studies of azimuthal metallicity distributions of the Milky Way \citep{Balser2011, Balser2015, DeCia2021, Pedicelli2009, Davies2009} and external spiral galaxies \citep{Ho2017,Ho2018,Kreckel2019,Kreckel2020}.  The result suggests that the ISM  is not well mixed at any time.

 We find five metal-poor substructures with sizes of $\sim$ 0.2--1.0\,kpc,  metallicities of which are smaller than their nearby regions by $\sim 0.1$\,dex.  These metal-poor substructures may be associated with high-velocity clouds infalling into the Galactic disk from the Galactic halo.    According to the results of stellar populations at different ages, we suggest that high-velocity clouds infalling into the Galactic disk from the Galactic halo are not well mixed with the pre-existing Galactic disk ISM within 0.3\,Gyr. 

The size and spatial distribution of our stellar sample are mainly limited by the faint limiting magnitude (15\,mag in $g$-band) of the LAMOST Medium-Resolution Spectroscopic Survey.   Future large-scale spectroscopic surveys, including  SDSS-V \citep{Kollmeier2017,Zari2021} and 4MOST \citep{deJong2022},  are expected to improve our work by enlarging the sample size and spatial coverage. 


\begin{acknowledgements}
 We appreciate the helpful comments of the anonymous referee. This work was funded by the National Key R\&D Program of China (No. 2019YFA0405500) and the National Natural Science Foundation of China (NSFC Grant No.12203037, No.12222301, No.12173007 and No.11973001).  We acknowledge the science research grants from the China Manned Space Project with NO. CMS-CSST-2021-B03.  We used data from the European Space Agency mission Gaia (http://www.cosmos.esa.int/gaia), processed by the Gaia Data Processing and Analysis Consortium (DPAC; see http://www.cosmos.esa.int/web/gaia/dpac/consortium). Guoshoujing Telescope (the Large Sky Area Multi-Object Fiber Spectroscopic Telescope LAMOST) is a National Major Scientific Project built by the Chinese Academy of Sciences. Funding for the project has been provided by the National Development and Reform Commission. LAMOST is operated and managed by the National Astronomical Observatories, Chinese Academy of Sciences.
\end{acknowledgements}

\bibliography{metallicity_variations}{}
\bibliographystyle{aa}


\begin{appendix}
\section{Additional figures}

\begin{figure*}
\centering
\includegraphics[width=6.5in]{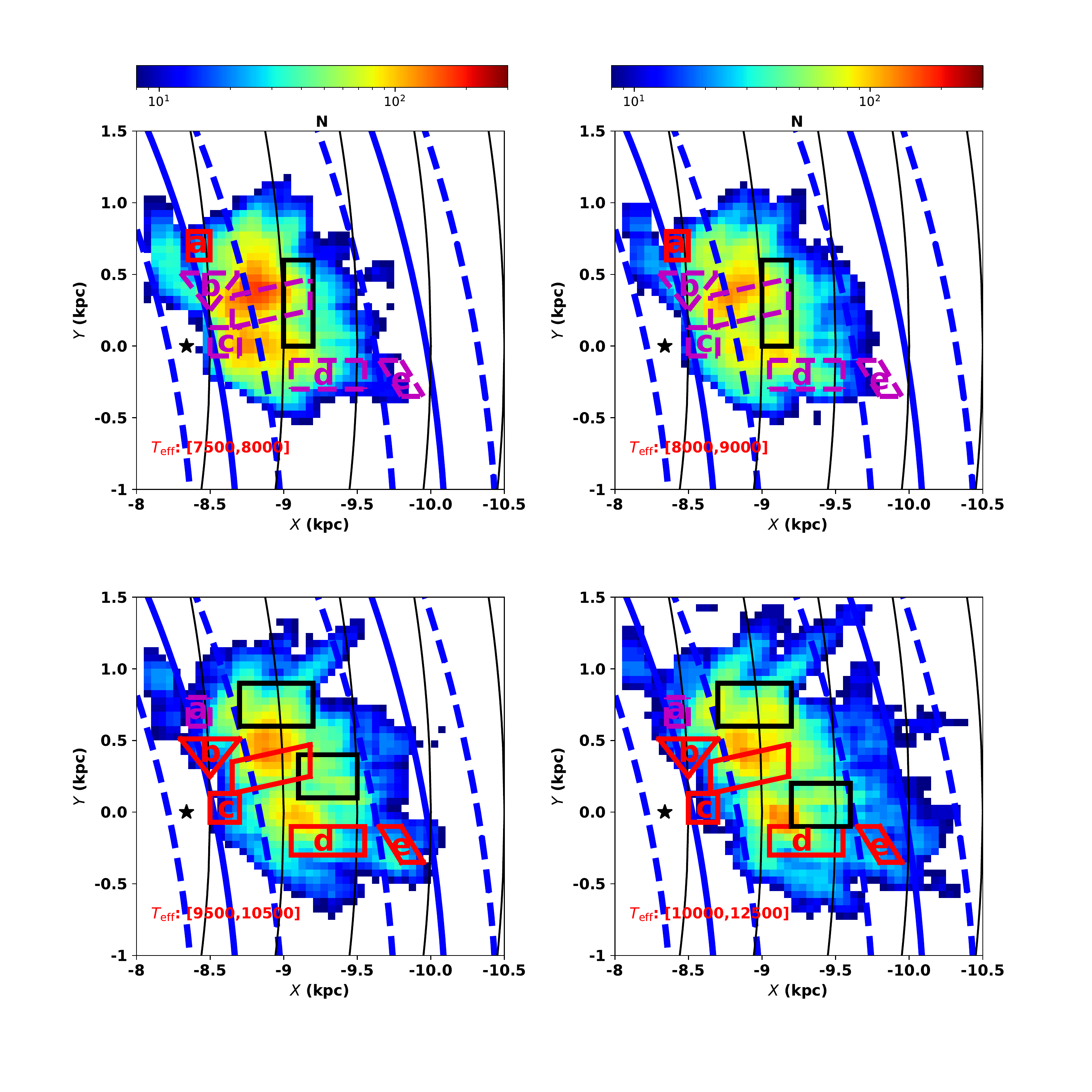}
\caption{Same to Figure\,\ref{metallicity_xy_all_bootstrap}, but for the stellar number distributions.}
\label{number_xy_all_bootstrap}
\end{figure*}

\begin{figure*}
\centering
\includegraphics[width=6.5in]{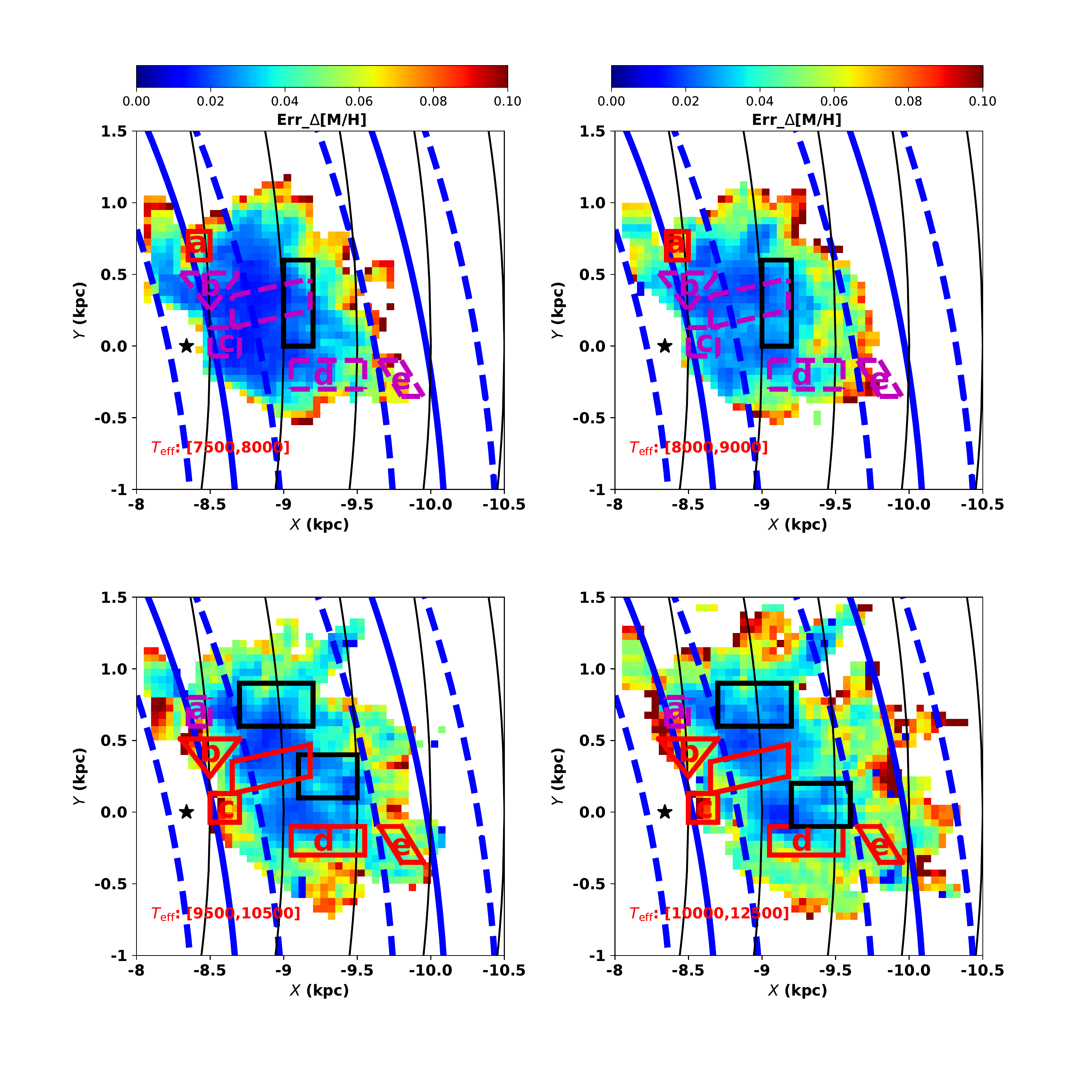}
\caption{Same to Figure\,\ref{metallicity_xy_all_bootstrap}, but for the distributions of metallicity excess uncertainties. }
\label{metallicity_xy_all_bootstrap_err}
\end{figure*}

\begin{figure*}
\centering
\includegraphics[width=6.5in]{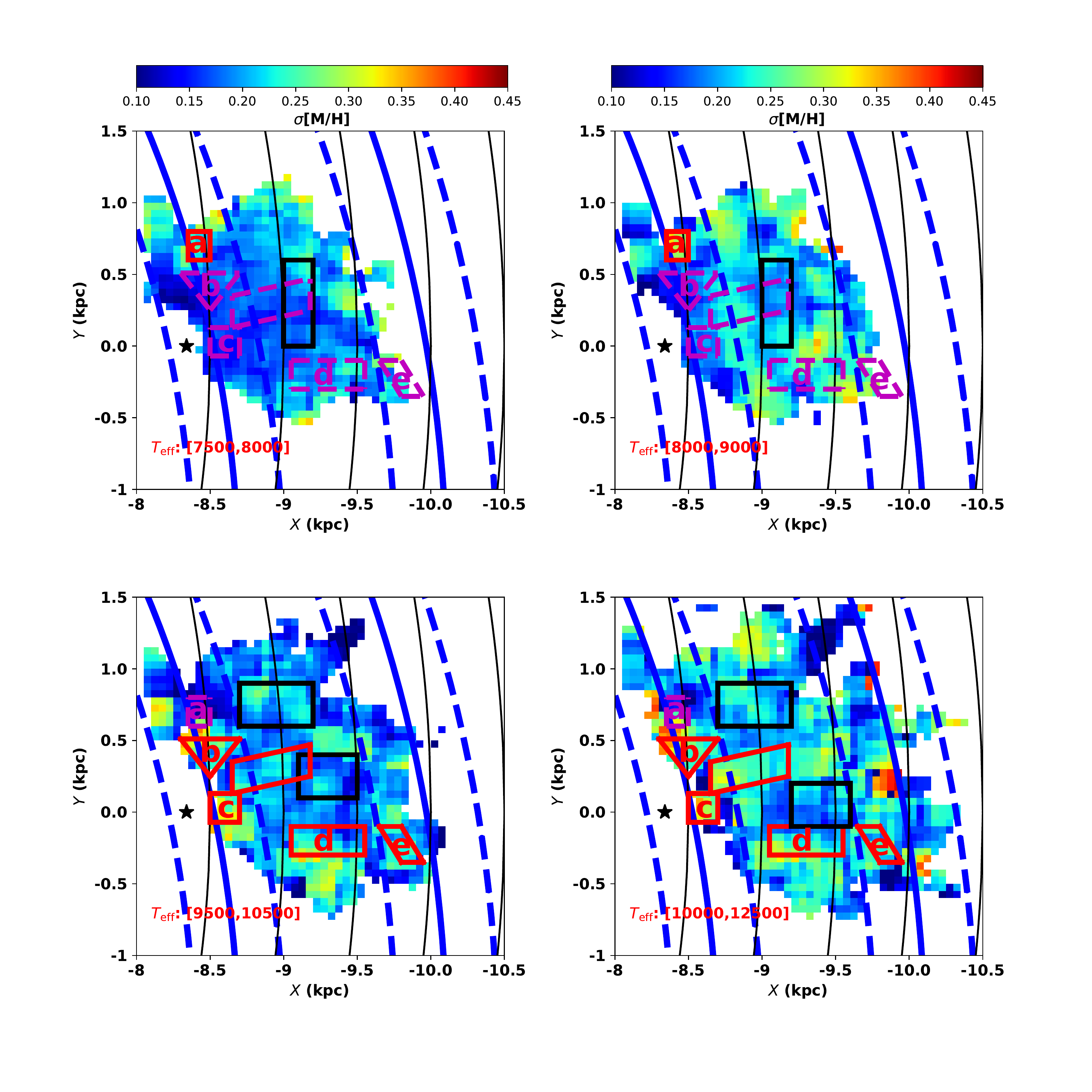}
\caption{Same to Figure\,\ref{metallicity_xy_all_bootstrap}, but for the distributions of the dispersion of the metallicity excess. }
\label{sigma_metallicity_xy_all}
\end{figure*}

\begin{figure*}
\centering
\includegraphics[width=6.5in]{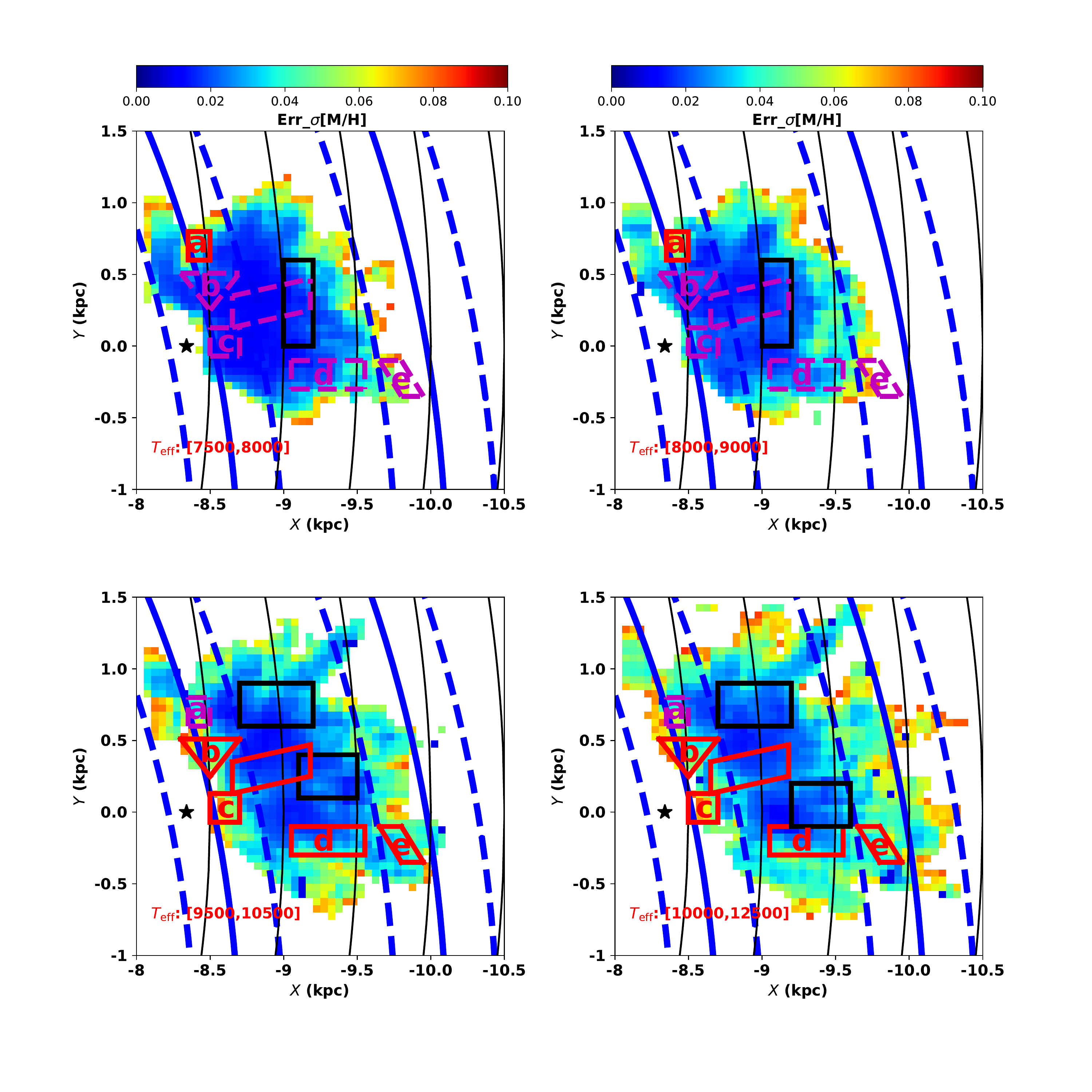}
\caption{Same to Figure\,\ref{metallicity_xy_all_bootstrap}, but for the distributions of metallicity excess dispersion's uncertainty. }
\label{sigma_metallicity_xy_all_err}
\end{figure*}

\begin{figure*}
\centering
\includegraphics[width=6.5in]{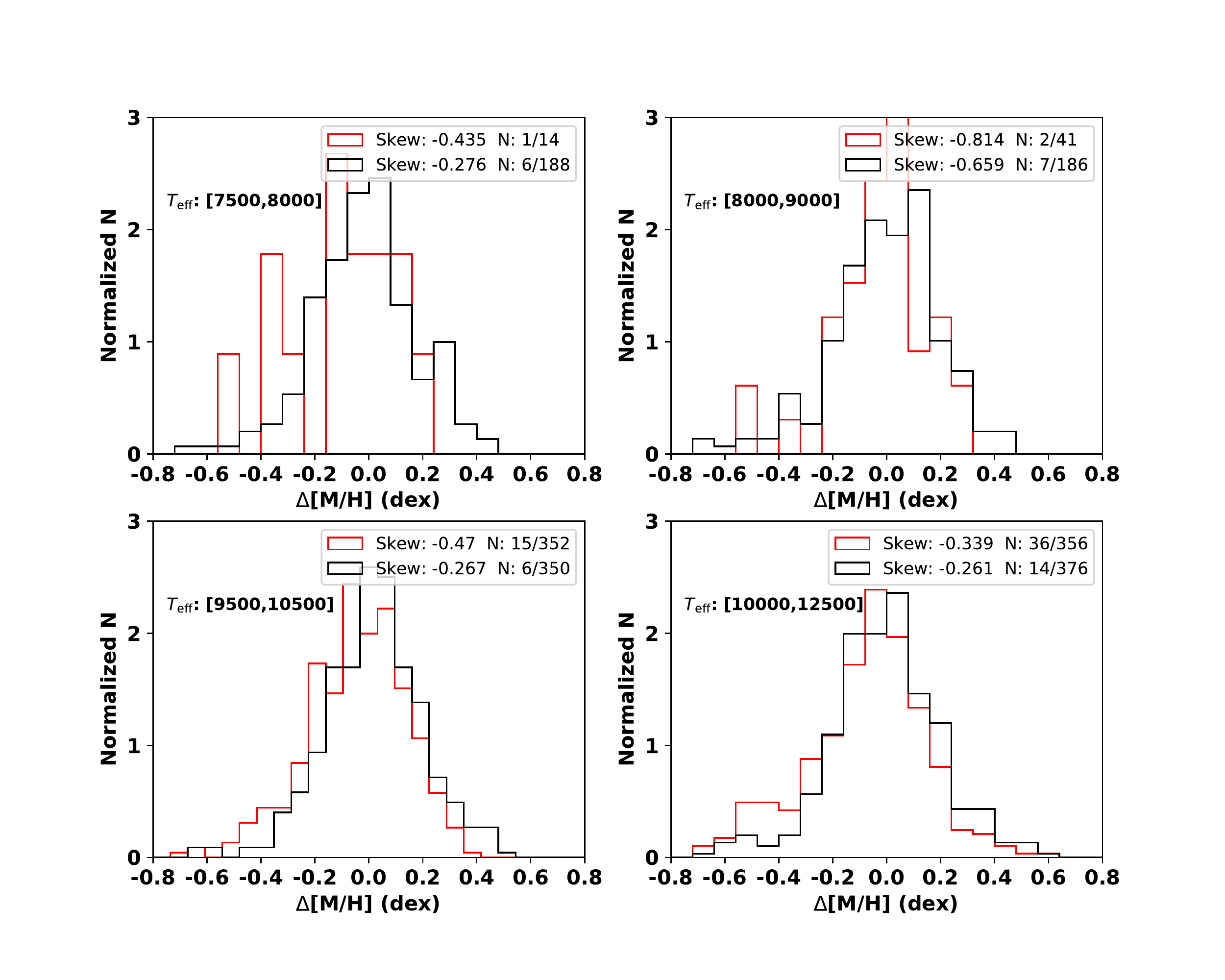}
\caption{Metallicity excess distributions in metal-poor regions (red histogram) and their control regions (black histogram) of these four mono-temperature stellar populations.  The skewness of these distributions is estimated and labelled at the top right corner of each panel.  The total number and the number of stars with $[\rm M/H]_{corr} < -0.4$\,dex are also shown in the figure. }
\label{metallicity_dispersion_compare}
\end{figure*}

\end{appendix}

\end{document}